\def\beq{\begin{equation}}

\def\bbb1{{\rm 1\!1}}

\newcommand{\refs}[1]{(\ref{#1})}

\documentclass[12pt]{article} 
 
\usepackage{epsfig} 
 
\def\>{\rangle} 
\def\<{\langle} 
\def\mtx#1{\quad\hbox{{#1}}\quad} 
 
\def\be{\beta}

\def\be{\beta}

\def\nn{\nonumber}

\usepackage{epsfig}
\usepackage{latexsym}
\usepackage{amssymb}

\def\be{\begin{equation}}
\def\ee{\end{equation}}
\def\ba{\begin{eqnarray}}
\def\ea{\end{eqnarray}}

\title{{\bf Tachyon Dynamics and the  Effective Action
Approximation}}

\author{\Large{N.D. Lambert}${}^1$ and \Large{I. Sachs}${}^2$\\ 
\\
${}^1$Dept. of Physics and Astronomy\\ 
Rutgers University\\ 
Piscataway, NJ 08855\\ 
USA\\ 
nlambert@physics.rutgers.edu\\ 
\\
${}^2$School of Mathematics\\ 
University of Dublin\\
 Trinity College\\
College Green, Dublin 2\\
Ireland\\ 
ivo@maths.tcd.ie\\
\\
RUNHETC-02-31\\
hep-th/0208217}

\begin{document}

\maketitle

\begin{abstract}
Recently effective actions have been extensively used 
to describe tachyon condensation in string theory. 
While the various effective actions which have appeared
in the literature have very similar properties for static configurations,
they differ for time-dependent tachyons. In this paper we 
discuss 
general properties of non-linear effective Lagrangians which are first 
order in
derivatives. In particular we show that some observed properties, such as 
asymptotically vanishing pressure, are rather generic features, although the
quantative features differ. On the other hand we argue that
certain features of marginal tachyon profiles are 
beyond the reach of any first order Lagrangian description.
We also point out that an effective action, proposed earlier,  
captures the dynamics of tachyons well.
\end{abstract}

{Keywords: String Theory, D-branes, Tachyons} 

\eject


\section{Introduction} 

The decay of unstable 
D-branes is an important and challenging problem in string theory. 
In order to understand such processes it is important to obtain a 
reliable description of 
the dynamics of the tachyons which arise from 
open strings on unstable D-branes. 
In the last few years considerable progress has been made 
starting with Sen's proposal to identify the closed string vacuum with the 
tachyon vacuum on unstable D-branes in superstring theory 
(see \cite{Sen:1999mg} for a review and references). 
Furthermore, a class of time independent kink solutions corresponding to lower 
dimensional D-branes was identified with marginal boundary 
deformations of the open string sigma model \cite{Sen:1999mg}. 
More recently  Sen has obtained a family of time dependent 
tachyon solutions as marginal deformations 
of the open string sigma model \cite{Sen:2002nu} 
(see also \cite{Gutperle:2002ai}). 
An analysis of  the 
stress tensor obtained from the boundary state for a decaying D-brane 
\cite{Sen:2002in} shows that the decay of an unstable D-brane results 
in a gas with finite energy density but vanishing pressure. 
 
An interesting question 
is then to what extent these features can be obtained from 
a first derivative effective action for the tachyon 
\cite{Minahan:2000ff,Minahan:2000tg,Sen:2002an}. 
Typically a tachyon effective action contains an infinite number 
of higher derivative terms  which, unlike the case for massless 
string modes, cannot  be  simply neglected. Therefore no truncation
to first derivatives can be expected to capture all of the dynamics
of the tachyon. 
Furthermore, even if the full effective action
were known, without some kind of simplifying structure 
it would be near impossible to obtain concrete 
results.\footnote{We note that \cite{MZ} recently presented a
very interesting analysis of tachyon dynamics in p-adic string
theory, where the effective action is known to all orders in
derivatives and is amenable to numerical analysis.} 
In addition it is not clear whether an 
initial value problem can be formulated. However first 
derivative 
truncations of the tachyon effective action have been rather sucessful  
for describing  D-branes as static tachyonic solitons 
\cite{Harvey:2000tv}-\cite{Takayanagi:2000rz}. One may therefore hope that 
a truncated action could also be useful to describe the dynamics of 
tachyon condensation. 

In \cite{Sen:2002an} Sen 
argued that some qualitative features of full tree-level string theory,
such as the asymptotic vanishing of the pressure, are 
indeed reproduced by a Born-Infeld (BI) type  action 
\cite{Garousi:2000tr}.  
Unlike the case of massless gauge fields, the BI-type action has not been 
inferred from on-shell string theory and marginal tachyon profiles do
not solve the equations of motion. On the other hand there is a well 
defined prescription to extract 
the tachyon effective action from boundary string field theory (BSFT) 
\cite{Witten:1992cr}-\cite{Shatashvili:1993ps}, 
\cite{Gerasimov:2000zp}-\cite{Takayanagi:2000rz}. Unfortunately this
action is known explictly only for constant and 
linear tachyon profiles and the marginal tachyon profiles are also  
not solutions of the BSFT effective action. Nevertheless BSFT has been 
quite sucessful in describing some aspects of tachyon condensation to 
lower dimensional D-branes 
\cite{Gerasimov:2000zp,Kutasov:2000qp,Kraus:2000nj}. Furthermore it was 
shown in \cite{0205085,0205098} that some qualitative features of the 
tachyon dynamics obtained from the BSFT-action are consistent with conformal 
field theory results. However, while the various effective actions that
have appeared in the literature are remarkably similar for static profiles, 
they could hardly differ more for time dependent ones.
This is illustrated in 
figure $1$ where the kinetic part of the effective action
is plotted as a function of $\partial_\mu T\partial^\mu T$ for three
different proposals (the details will be explained below).

\begin {figure}
\center{\epsfig{file=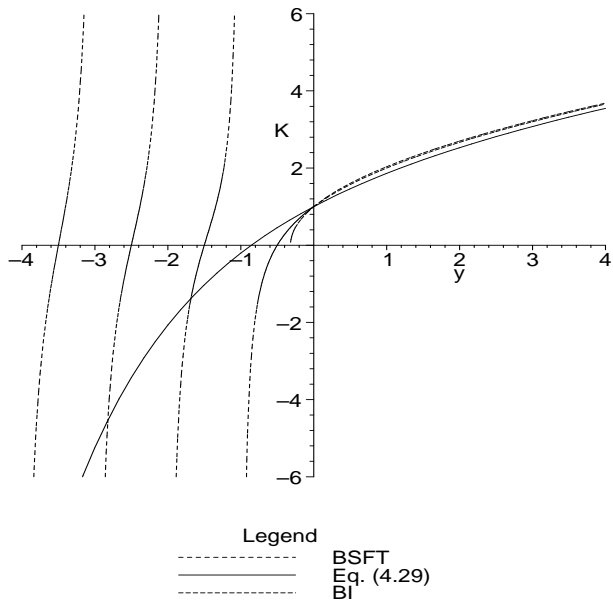,height=8cm,width=8cm}}
\caption{$K(y)$ from various effective actions.}
\end{figure}

Given this state of affairs, it is of interest establish to what extent 
these properties are generic in non-linear, first derivative effective 
actions for the tachyon. In addition we would like to determine the
most appropriate behaviour of the effective action for time dependent
configurations.
The purpose of this paper is establish general properties of suitable 
first derivative tachyon actions derived under a minimal set of reasonable 
theoretical assumptions. 
We find that some observed qualitative phenomena in tachyon 
condensation are rather generic and are reproduced by a large class of 
effective actions, although the exact quantitative predictions differ
significantly.  
On the other hand we find that some global properties 
of marginal tachyon profiles are beyond the scope of first derivative 
effective actions. 

In this context we also analyse a tachyon action proposed earlier 
\cite{Lambert:2001fa} 
which closely resembles the other effective 
actions for time independent configurations, 
but has the advantage that marginal deformations are solutions 
to the equations of motion. Furthermore, the time-dependent marginal 
tachyon profiles in 
\cite{Sen:2002nu} are also exact classical solutions for this 
action. We find that equation of state obtained from this action is  
in good quantitative agreement with the exact results of conformal field
theory. Finally the potential cosmological 
relevance of scalar field actions with non-standard kinetic terms has been 
pointed 
out in \cite{Armendariz-Picon:1999rj}. It 
is therefore 
interesting to analyse tachyon effective actions from this point of view. 
This was 
done in \cite{Gibbons} for the BI-type action and in \cite{0205085} for 
the BSFT-action. We will see that cosmological implications of our 
tachyon action are qualitatively similar to these results, although the 
details are again different. 
  
\section{Tachyons as Marginal Deformations } 

Finding the general tachyon solution in string theory 
would require a detailed knowledge of open superstring field theory. 
However, 
using the fact that these solutions correspond to conformally invariant 
backgrounds for the open string sigma model, a subset of open string 
tachyon 
solutions on a non-BPS $D$-brane in type II string theory can be 
obtained as exactly marginal 
deformations of the boundary conformal field theory. 
Specifically, as is well known 
\cite{Callan:1994ub}-\cite{Lambert:2000cg}, 
\begin{equation}\label{marginal}
T(x^p) = \chi {\rm sin}(x^p/\sqrt{2\alpha'})\ \mtx{and}\  T(x^p) = \chi {\rm cos}(x^p/\sqrt{2\alpha'})
\end{equation}
are exactly marginal 
deformations of the open string Polyakov action on a non-BPS $Dp$-brane. The 
physical properties of this open string background are encoded in the 
corresponding boundary state $|B,+\>-|B,->$ \cite{POLCAI,Sen:2002in}, 
where\footnote{Our metic convention is $(-,+,\cdots,+)$
and $\mu,\nu=0,...,p$ label the non-BPS D$p$-brane worldvolume.}  
\begin{equation}
|B,\epsilon>=|B,\epsilon>_{x^p,\psi^p}\otimes |B,\epsilon>_{x^\mu,\psi^\mu}\otimes |B,\epsilon>_{ghosts}\ ,\qquad \mu\neq p\ ,\qquad \epsilon=\pm\ .
\end{equation}
Here $|B>_{x^\mu,\psi^\mu}$ and $|B>_{ghosts}$ are the same as for a $p-1$ 
brane, while $|B>_{x^p,\psi^p}$ describes the $x^p$ dependence of the 
marginal deformation. 
In particular, the energy momentum tensor for a marginal  kink profile  
can be obtained by evaluating the matrix element of the boundary state 
$|B>$ with the closed string graviton state $<g_{\mu\nu}|$. 
Concretely we have
\begin{equation}\label{stress}   
T_{\mu\nu}(x)=-\frac{\tilde\tau_p}{2}
\left(A_{\mu\nu}(x)+B(x)\eta_{\mu\nu}\right) \ . 
\end{equation} 
Here $\tilde\tau_p=\sqrt{2}\tau_p$ is the tension of a non-BPS $D$-brane 
while $A_{\mu\nu}$ and $ B$ parametrise the level 
$(\frac{1}{2},\frac{1}{2})$ 
part of the boundary state \cite{Sen:2002in} 
\begin{equation}
|B>_{(\frac{1}{2},\frac{1}{2})}=\sum\limits_{k_p}\left((\tilde 
A_{\mu\nu}(k_p)\psi^\mu_{-1/2}\psi^\nu_{-1/2}+\tilde 
B(k_p)(\bar\beta_{-\frac{1}{2}}\gamma_{-\frac{1}{2}}-\beta_{-\frac{1}{2}}\bar 
\gamma_{-\frac{1}{2}})\right)|\Omega, k>\ ,
\end{equation}
with
\begin{equation}
|\Omega, k>=(c_0+\bar c_0)c_1\bar c_1 e^{-\phi(0)}e^{-\bar\phi(0)}|k>
\end{equation}
and $|k>$ is the closed string Fock vacuum with momentum $k$. 
Using similar arguments as in \cite{Sen:2002in} we then find,
for a $\cos(x^p/\sqrt{2\alpha'})$ kink,
\begin{eqnarray}
A_{\mu\nu}(x)&=&-f(x^p)\eta_{\mu\nu}\ ,\quad \mu,\nu\neq p\nn\\
A_{pp}(x)&=&-g(x^p)\ ,\,\qquad B(x)=-f(x^p)\ ,
\end{eqnarray}
where
\begin{eqnarray}
f(x^p)&=&1+2\sum\limits_{n=1}^{\infty}\left(-\sin^2(\chi\pi)\right)^n 
\cos(n\sqrt{\frac{2}{\alpha'}}x^p)\nn\\
&=&\frac{1-\sin^4(\chi\pi)}{
1+2\sin^2(\chi\pi)\cos(\sqrt{\frac{2}{\alpha'}}x^p)+\sin^4(\chi\pi)}\ ,
\end{eqnarray}
and 
\begin{equation}
g(x^p)=2-2\sin^2(\pi\chi)-f(x^p)\ .
\end{equation}
In particular $T_{pp}$ is independent of $x^p$ as it should be. For $\chi$ 
close to $1/2$ the energy density is peaked about 
$x^p=\sqrt{\alpha'}(2n+1)\pi/\sqrt{2}$. 

We note that the boundary 
state and, in particular all components of the stress tensor are 
periodic in $\chi$ with period $1$. This periodicity can also bee seen by 
analysing the spectrum \cite{Sen:1998tt} or by analysing the cylinder diagram 
on an orbifold \cite{Lambert:2000fn}. The  origin of this
periodicity can be traced to the fact that the vertex operators 
${\rm exp}({ix^p}/{\sqrt{2\alpha'}}), 
{\rm exp}(-{ix^p}/{\sqrt{2\alpha'}})$ and 
$i\partial x^p$ form an $ {so}(3)$ current alegbra. 
Here $i\partial x^p$ generates translations along the $x^p$ direction.
Since the exponential of these operators appears in the path integral, the
resulting correlation functions gain a periodic dependence on
the coefficient $\chi$ due to the compactness of $so(3)$. 

Time dependent boundary perturbations can be obtained by analytic 
continuation of the above profile \cite{Sen:2002in}. 
Concretely, by performing the double Wick 
rotation $x^p\to ix^0, x^0\to -ix^p$ we 
find that $T = \chi {\rm sinh}(x^0/\sqrt{2\alpha'})$ and 
$T = \chi {\rm cosh}(x^0/\sqrt{2\alpha'})$ are, at  tree level, 
exactly marginal, time dependent tachyon profiles. Note that in the
${\rm sinh}(x^0/\sqrt{2\alpha'})$ case 
we must also Wick rotate $\chi \to -i\chi$ to
obtain a real profile. Therefore we expect that the peridocity
observed for $\chi$ in the space-like case is now broken. 
Indeed for space-like marginal deformations we saw that the 
perodicity in $\chi$ came from an $ {so}(3)$ current
alegbra of the $x^p$ free Boson CFT. If we Wick rotate to a
time-like free Boson this becomes an $ {so}(2,1)$
current algebra, which has only one compact direction. This must
correspond to the deformations generated by 
$\chi{\rm cosh}(x^0/\sqrt{2\alpha'})$
since these arise from Wick rotating 
$\chi{\rm cos}(x^p/\sqrt{2\alpha'})$ which does not 
require that $\chi$ is also Wick rotated.
In addition $i\partial x^0$ now generates time translations 
which are  no longer periodic.

The stress tensor for the time-like $\chi{\rm cosh}(x^0/\sqrt{2\alpha'})$ 
profiles is  again given by \refs{stress} with 
\cite{Sen:2002in} 
\begin{equation} 
A_{00}(x)=2-2\sin^2(\pi\chi)-f(x^0)\ ,\qquad A_{ij}(x)=-f(x^0)\delta_{ij}\ 
,\qquad  B(x)=-f(x^0)\ , 
\end{equation} 
and 
\begin{eqnarray}\label{fg} 
f(x^0)&=&\frac{1-\sin^4(\chi\pi)}{
1+2\sin^2(\chi\pi)\cosh(\sqrt{\frac{2}{\alpha'}}x^0)+\sin^4(\chi\pi)}\ .
\end{eqnarray} 
For time dependent marginal deformations the conserved 
energy $E=-T_{00}$ is
\begin{equation}
E=\sqrt{2}\tau_p(1-\sin^2(\chi\pi))\ ,
\label{Econ}
\end{equation}
while the 
pressure $p=-T_{ii}$ vanishes  as the tachyon approaches 
the vacuum configuration. More specifically, as $x^0\to \infty$, 
\begin{equation}\label{p} 
  p\simeq 
-E\frac{1+\sin^2(\chi\pi)}{\sin^2(\chi\pi)}e^{-\sqrt{\frac{2}{\alpha'}}x^0 }\ .
\end{equation} 
If we consider marginal deformations of
the form $\chi{\rm sinh}(x^0/\sqrt{2\alpha'})$ then we find similar
expressions but with $\sin^2(\chi\pi)$ replaced by 
$-{\rm sinh}^2(\chi\pi)$ and 
$e^{\pm\sqrt{\frac{2}{\alpha'}}x^0}$ replaced by 
$-e^{\pm\sqrt{\frac{2}{\alpha'}}x^0}$
\cite{Sen:2002in}.

\section{Tachyon Effective Actions} 

In the previous section we saw that 
in principle the physical observables of a marginal tachyon profile 
are encoded in the boundary state. An interesting question is then 
to what extent  can these properties be reproduced 
by an effective field theory for the tachyon. A reliable effective
action formulation would facilitate more complicated
calculations, such as the dynamics in the presence of evolving
closed string modes. For example in 
\cite{Sen:2002an} Sen showed 
that for time dependent tachyon solutions in a BI-type action the 
pressure does indeed vanish asymptotically. However, as we remarked in
the introduction, this particular form for the effective action is
not derived from string theory and one might therefore question the 
predictions obtained from it. In view of this we 
consider in this section the predictions of a rather
general class of effective actions. 

Typically, the full tachyon effective action will be non-local and thus 
difficult to handle. In what follows will consider effective actions which 
involve only first derivatives of the tachyon field. This class of actions 
includes, in particular, all suggestions which have appeared in the 
literature 
so far.
For a non-BPS brane it is known that the action and in particular
the potential $V$ must
be an even function of the tachyon field  $T$.
Therefore the most general form for the effective 
action of a real $T$ in $p+1$ dimensions that 
depends on at most first order derivatives and is even in $T$ is 
given by 
\begin{eqnarray} \label{action} 
S = \int d^{p+1} x {\cal L} 
\equiv \int d^{p+1} x \sum_{\alpha, \beta=0}^{\infty} 
c_{\alpha\beta}T^{2\alpha} 
(\partial_\mu T\partial^\mu T)^\beta\ . 
\end{eqnarray} 
From Noether's theorem it  folows that for any static, codimension one 
solution of the equations of motion of an action of the type 
\refs{action}  will have the integral of motion
\begin{equation} \label{kinkeq} 
T_{pp}={\delta {{\cal L}}\over \delta T'}T'-{\cal L}  = V_0\ , 
\end{equation} 
where $V_0$ is a constant and $T$ depends only on $x^p$. In addition the 
energy density $T_{00}={\cal L}$ is trivially 
conserved. For a time dependent but spatially homogeneous solutions
one again finds two conserved quantities but now the conservation
of energy leads to a non-trivial condition
 \begin{equation} \label{kinkt} 
T_{00}=-{\delta {{\cal L}}\over \delta \dot T}\dot T+{\cal L}  =
-E_0\ , 
\end{equation}
where $E_0$ is a constant.

The simplest condition one may want to impose is that the effective action
reproduces the correct perturbative tachyon mass near $T=0$, that is
${\cal L} \sim -(\partial T)^2+\frac{1}{2\alpha'}T^2+\ldots$. However
this also illustrates one problem of trunctating to first derivative
actions since integration by parts allows us to write the kinetic term 
as $T\partial^2 T$ and it could therefore be modified by higher derivative
terms. Indeed several proposals for
the effective action do not reproduce the correct tachyon mass.

The next issue that we want to discuss is whether the marginal deformations in 
the last section can be solutions of a first derivative effective action. 
As shown in \cite{Lambert:2001fa}, the requirement that 
$T = \chi {\rm sin}(x^p/\sqrt{2\alpha'})$ solves the field equations 
completely determines the general action 
\refs{action} in terms of an arbitrary potential 
$V(T)=f(\frac{T^2}{2\alpha'})$, i.e. 
\begin{equation}\label{KV} 
{\cal L} = \sum_{\gamma=0}^{\infty}{1\over \gamma !}{1\over 2\gamma-1} 
{d^\gamma f(t)\over dt^\gamma}(\partial_\mu T\partial^\mu T)^\gamma\ , 
\end{equation} 
where $t\equiv  T^2/{2\alpha'}$. 
Evaluating the resulting kink equation \refs{kinkeq} one then finds 
\begin{eqnarray}\label{sugg} 
V_0={\delta {\cal L}\over \delta T'}T'- {\cal L} 
&=& \sum_{\gamma=0}^{\infty}{1\over \gamma !} 
{d^\gamma f(t)\over dt^\gamma}(  (T')^2)^\gamma\cr 
&=& f\left( \frac{T^2}{ 2\alpha'}+(T')^2\right)\ . 
\label{kinkagain} 
\end{eqnarray} 
Thus, assuming that $V$ is nowhere constant, we see that 
the only regular static solutions are 
\begin{equation} 
T = \chi {\rm sin}\left({x^p-x_0\over\sqrt{2\alpha'}}\right)\ , 
\label{tkink} 
\end{equation} 
for arbitrary $x_0$ and $\chi$. In addition 
it is easy to see that, by taking the limit
$\chi\to 0$, this condition also ensures that the correct perturbative
mass for the tacyhon is reproduced.

Let us now look at the periodicity properties of the physical observables 
on marginal profiles. As discussed in section $2$ the stress tensor for 
marginal deformations is periodic in $\chi$. For a space-like kink 
$T = \chi {\rm sin}({x^p}/{\sqrt{2\alpha'}})$, 
periodicity of the conserved energy $T_{00}$ in $\chi$ 
implies that ${\cal L}$ is periodic in $\chi$ for all values in $x^p$. 
This implies that ${\cal L}(T,T')=F(\frac{T^2}{2\alpha'}+T'^2)$, where
$F$ is a periodic function. 
On the other hand periodicity of the conserved momentum $T_{pp}$ implies that 
\begin{equation} 
\frac{\delta{\cal L}}{\delta T'}T'- {\cal L}
=2\frac{d F(z)}{d z}T'^2- F(z)
\end{equation} 
is periodic in $\chi$. These two conditions are, however, incompatible. 
Similar comments apply to the case of time-dependent solutions.
Thus, we conclude that while marginal deformations can be solutions 
of first derivative effective actions, the periodicity in $\chi$ of the 
all observables cannot be reproduced by any first derivative 
effective action. This problem, which  was also observed 
in \cite{Lambert:2000fn},
originates in the fact that the string ground state at
$\chi=0$ is not the same as the string ground state in the new vacuum at
$\chi=1/2$ due to spectral flow. Therefore to see this periodicity the 
infinite tower of massive strings modes has to be taken into account. 
Generally we then expect that while effective actions capture the physics 
near a given string theory vacuum, global properties 
are typically beyond such approximations. In particular we then 
expect that the effective action \refs{KV} will only be reliable 
for $\chi<<1$.

In superstring theory all of the
tachyon effective actions proposed to date take the specific 
form\footnote{Although we note that for the Bosonic string the
action does not factorise 
\cite{Gerasimov:2000zp,Kutasov:2000qp,Lambert:2001fa}.} 
\begin{equation} \label{ex1} 
{{\cal L}} = -V(T)K(\partial_\mu T\partial^\mu T)\ . 
\end{equation} 
Certainly without some kind of simplifying structure
even an action which has been truncated to be only 
first order in derivatives becomes intractable.
Therefore in what follows we restrict our attention to
these forms for $\cal L$.  We note here that an effective
action of this form is compatible with T-duality if the transverse
scalars and worldvolume vector fields are included as
${\cal L} = -V(T)\sqrt{-\det(G_{\mu\nu}+{\cal F}_{\mu\nu})}
K((G+{\cal F})^{\mu\nu}\partial_{\mu} T\partial_{\nu}T)$ or in the BI form ${\cal L} 
= -V(T)\sqrt{-\det(G_{\mu\nu}+{\cal F}_{\mu\nu}+\kappa_{BI}\partial_\mu T\partial_\nu T)}$ \cite{Kamimura:2000bi,Bergshoeff:2000dq}.

Sen's conjectures state
that the potential is nowhere negative, vanishes
in the true open string vacuum and has a maximum value of
$\sqrt{2}\tau_p$ at $T=0$.  
Furthermore, there should be no perturbative
open string excitations about the true vacuum. Although one might think that
action of the form \refs{ex1} would ensure this condition, this is not
true in general. Assuming that for small 
$y$,  $K(y)\sim 1+\kappa_{1} y +...$ we 
check the perturbative excitations around $V=0$ by 
introducing a new tachyon variable.
\begin{equation}
\label{phidef}
d\varphi = \sqrt{V(T)}dT\ .
\end{equation}
The effective action now 
looks like 
\begin{equation}\label{Lphi} 
{\cal L} = -V(\varphi)K\left({\partial_\mu\varphi\partial^\mu\varphi\over 
V}\right)\ . 
\end{equation} 
Note that this changes of variable relates the BI action \cite{Garousi:2000tr} to the proposal of
\cite{Kluson:2000iy}.
Expanding \refs{Lphi} for $(\partial \varphi)^2<< V$ ({\it i.e.}
$(\partial T)^2<<1$) leads to 
\begin{equation}
{\cal L} \simeq -\kappa_{1}\partial_\mu\varphi\partial^\mu\varphi-
V(\varphi)\ .
\end{equation}
The mass-squared 
of elementary excitations  around the tachyon vacuum is then given 
by
\begin{equation}\label{mass}
M^2=\frac{2}{\kappa_{1}}\frac{d^2V(\varphi)}{d\varphi^2}\large|_{V=0}
=\frac{2}{\kappa_{1}}\frac{1}{\sqrt{V}}{d^2\sqrt{V}\over
  dT^2}
\large|_{V=0}\ .
\end{equation}
The absence of perturbative excitations implies that this mass is
infinite. It is not hard to convince oneself that $M$ will be  
infinite provided that $V$ vanishes faster than 
$V \sim e^{-a T}$ if the minimum is at
infinite $T$, or faster than 
$V\sim (T-T_0)^2$ if the minimum is at a finite
value of $T$. 
Note that $V \sim e^{-a T}$ 
appears in the effective action of a D-brane in {\it Bosonic} string theory.
However the effective actions of Bosonic  BSFT 
\cite{Gerasimov:2000zp,Kutasov:2000qp,Lambert:2001fa} 
do not have the form \refs{ex1} and hence \refs{mass} does not
apply. 
Indeed one can check that the resulting mass in these cases is
infinite. 
Nevertheless, even if the form \refs{ex1} is assumed, 
plane-wave excitations for $\varphi$ are absent due to  
non-linearities in the kinetic term \cite{Sen:2002an}. 
It has also been pointed out that the Hamiltonian
formalism is in fact better suited for analysing the dynamics
in the true vacuum \cite{HGY}. However this cannot be done here without
choosing a specific form for $K$.

Another requirement that the effective action should satisfy is
that it should reproduce the correct tension for kink solutions
which are identified with a BPS D$(p-1)$-brane.
More generally  if $T = uf(x^p)$ 
is an off-shell profile then $u\to \infty$ 
in the infra-red limit.  The worldsheet theory
is expected to run to that of $N$ BPS D$(p-1)$-branes and
anti-D$(p-1)$-branes. Here
$N$ is the number of times $T$ interpolates between the vacua as
$u\to \infty$. Since this should be true
for all profiles this suggest that at large $u$, {\it i.e.} at large
$\partial_p T$, the action becomes topological. This will
be the case if $K \simeq \kappa_{\infty}\sqrt{(\partial T)^2}$ as
$(\partial T)^2\to +\infty$. Recall
that we normalise $K(0)=1$ and 
$V(0)=\sqrt{2}\tau_p$ so that $\kappa_{\infty}$ is not
arbitrary. In this case the energy for 
the profile $T = uf(x^p)$ is, in the limit $u\to \infty$,
\begin{equation}\label{tension}
E = \kappa_{\infty} \int_{-\infty}^{\infty} dx^p V(uf(x^p)) u|f'(x^p)|
= N \kappa_{\infty} \int  dT\ V(T)\ .
\end{equation}
It is not hard to see that this property is indeed shared 
by all non-linear, first derivative superstring tachyon 
actions proposed thus far \cite{Garousi:2000tr,Kutasov:2000aq,Kraus:2000nj,
Takayanagi:2000rz, Lambert:2001fa}. 
It was also pointed out in \cite{Minahan:2000tg} 
that a square-root form for time independent configurations is needed to ensure
that the fluctuations about a kink background have finite masses.
A square-root form also ensures that the effective action for the relative
centre of mass coordinates of the D$(p-1)$-branes 
has a BI form \cite{Lambert:2001fa}. In addition, 
to agree with the interpretation
as $N$ seperate BPS D$(p-1)$-branes and anti-D$(p-1)$-branes 
this energy should be  $E = 2\pi\sqrt{\alpha'}N$. Hence we find
a constraint on the area of the potential between two minima
and the large $(\partial T)^2$ behaviour of $K$. This condition is
indeed satisfied by the proposals of
\cite{Kutasov:2000aq,Kraus:2000nj,Takayanagi:2000rz,Lambert:2001fa}.

Lastly we consider  the asymptotic form 
of the pressure for homogenous, time-dependent tachyon approaching the 
minimum of the potential. We then have for the conserved energy
\begin{equation} \label{E}
E=-T_{00}=2V(T)K'\dot T^2+V(T)K((\partial T)^2)\ ,
\end{equation} 
while the pressure is given by 
\begin{equation} \label{pgen}
p=-T_{ii}=-V(T)K((\partial T)^2)\ .
\end{equation} 
Because $V(T)$ vanishes at its minimum, energy 
conservation implies that $\dot T^2$ necessarily approaches 
a singular point of $K'$ or $K$ as $T$ condenses to $V=0$. Now if this 
singularity is at some finite value of $\dot T$ then, 
since $K'$ will diverge faster than $K$, 
the first term in \refs{E} will dominate and the second term
will vanish. Therefore $p$ must also 
approach zero at this point. If the singular point of $K$ or $K'$ 
is at infinity then $p$ vanishes unless $yK'(y)$ and $K(y)$ have the 
same asymptotic behaviour; that is if $K(y)\sim y^n$ as 
$y=-\dot T^2 \to -\infty$.  In this latter case, as $V\to 0$, 
$p \to -E/(2n+1)$ is non-vanishing. Thus, unless $K(y)$ has a power law
behaviour  for  rapidly varying time dependent 
tachyons, the pressure vanishes as the tachyon condenses. 
Although whether or not this happens 
exponentially quickly or not will depend on the choice of $K$.

\section{Tachyon Dynamics} 

In the previous section we outlined various properties that the
effective action for the tachyonic mode of a non-BPS D-brane is 
expected to have. In particular
if we require that the marginal deformations \refs{marginal} 
are solutions to the equations of motion then the effective action 
is uniquely determined by a choice
of potential $V(T)$. 
As proposed in \cite{Lambert:2001fa}, to fix this ambiguity we take 
the exact potential found in boundary string field theory 
\cite{Kutasov:2000aq,Kraus:2000nj,Takayanagi:2000rz} 
\begin{equation}\label{VBSFT} 
V(T) = \sqrt{2}\tau_pe^{-{ T^2\over 2\alpha'}}\ , 
\end{equation} 
where $\tau_p$ is the tension of a BPS $Dp$-brane. 
The resulting Lagrangian that we construct then takes the 
form \cite{Lambert:2001fa} 
\begin{eqnarray}\label{kinkyaction} 
{\cal L} &=& -\sqrt{2}\tau_pe^{-{T^2\over 2\alpha'}} 
\left[e^{ - \partial_\mu T\partial^\mu T} 
+ \sqrt{\pi \partial_\mu T\partial^\mu T} 
{\rm erf}\left(\sqrt{ \partial_\mu T\partial^\mu T}\right) 
\right] 
\ .
\end{eqnarray} 
One can then check that this action satisfies all the 
properties discussed in the previous section (with the exception
of periodicity which we argued could not be captured by any
first derivative effective action).

For static configurations this action is in remarkable agreement
with the BSFT action \cite{Kutasov:2000aq,Kraus:2000nj,Takayanagi:2000rz} 
\begin{equation}
{\cal L} = 
\frac{1}{\sqrt{2}}\tau_p e^{-T^2/2\alpha'}4^{(\partial T)^2}(\partial T)^2
{\Gamma((\partial T)^2)^2 \over \Gamma(2(\partial T)^2)}\ .
\end{equation}
This is somewhat surprising since the BSFT action is
derived by simply assuming a linear tachyon profile. 
This good agreement
can be viewed as a test of the BSFT action for non-trivial space-like
kinks, since these are solutions to the equations of motion of
\refs{kinkyaction}. 
However, for time-like solutions the two actions differ considerably.
In particular, while \refs{kinkyaction} is smooth for all $(\partial
T)^2$
the BSFT action has poles at negative integer values of $(\partial T)^2$.
However there is no reason to believe that these 
poles are physically important since the BSFT action is derived for linear
profiles but these are not solutions to the equations of
motion. Furthermore, the fact that the action \refs{kinkyaction}
and the BSFT action differ substantially for time dependent solutions
suggests that, in contrast to space-like marginal profiles, the BSFT
effective action cannot be trusted. The action \refs{kinkyaction} also
agrees well with the BI form \cite{Garousi:2000tr}
\begin{equation}
{\cal L}=\sqrt{2}\tau_p e^{-T^2/2\alpha'}\sqrt{1+\kappa_{BI}(\partial T)^2}\ ,
\end{equation}
for $(\partial T)^2 > 0$ if we take $\kappa_{BI}=\pi$. 
However again they differ substantially for
time dependent profiles where $(\partial T)^2<0$ and therefore similar
comments apply. In particular
the BI form imposes a maximum value of $|\dot T|$. These three
forms for the function $K$ are plotted in figure $1$.

From the construction of the effective action in the last section it is 
clear that 
\begin{equation}\label{Tsol} 
T(x^0)=A\sinh(\frac{x^0}{\sqrt{2\alpha'}})+B\cosh(\frac{x^0}{\sqrt{2\alpha'}}) 
\end{equation} 
is an exact solution of the equation of motion. In fact we can say
more by analysing the energy momentum tensor for the action \refs{kinkyaction} 
\begin{equation}\label{tmunu} 
T_{\mu\nu}=-{\sqrt{2}}\tau_p e^{-{T^2\over 
2\alpha'}}\left[\frac{\sqrt{\pi}}{\sqrt{(\partial T)^2}}\partial_\mu 
T\partial_\nu T{\rm erf}\left(\sqrt{ (\partial T)^2 }\right)
-\eta_{\mu\nu}\left(e^{ - (\partial T)^2} 
+ \sqrt{\pi (\partial T)^2} 
{\rm erf}\left(\sqrt{ (\partial T)^2}\right)\right)\right] \ . 
\end{equation} 
Let us now consider a homogenous, but otherwise arbitrary, time dependent 
tachyon configuration. Then the energy takes the simple 
form 
\begin{equation}\label{energy} 
E=-T_{00}=\sqrt{2}\tau_pe^{-{ T^2\over 2\alpha'} + \dot T^2}
\ .
\end{equation}  
Conservation of energy then implies that \refs{Tsol} is the only regular 
solution of the equation of motion. Of course, 
the same result can be obtained by analytic continuation from \refs{sugg}. 
In particular, as the tachyon rolls to the minimum $\dot T$ diverges in 
agreement with the conformal field theory 
approach. This is in contrast  
with the BI-type and BSFT actions where $\dot T$ approaches a constant 
\cite{Sen:2002in,0205085,0205098}.
 
Let us now consider $T_{ij}$. From \refs{tmunu} we 
have 
\begin{equation}\label{pressure} 
T_{ij}={\sqrt{2}}\delta_{ij}
\tau_pe^{-{ T^2\over 2\alpha'}}\left(e^{\dot T^2} +i 
\sqrt{\pi\dot T^2}{\rm erf}\left(i\sqrt{\dot T^2}\right)\right)\ . 
\end{equation} 
Now, for large $y$, 
\begin{equation}\label{erf} 
\sqrt{\pi}{\rm erf}(iy) \simeq 
i\;e^{y^2}\left(\frac{1}{y}+ \frac{1}{2y^3}
+O\left(\frac{1}{y^5}\right)\right)\ , 
\end{equation} 
so that 
\begin{equation} \label{papprox}
T_{ij}\simeq
{T_{00}\over 2\dot T^2} \delta_{ij}
\mtx{ for} \dot T\to\infty\ .
\end{equation} 
Thus, the action \refs{kinkyaction} predicts that at large times 
the tachyon condensation produces a gas with non vanishing energy and 
vanishing pressure. In particular  for large $x^0$, where $T \simeq
\chi e^{x^0/\sqrt{2\alpha'}}/2$, 
\begin{equation}\label{asym} 
  p\simeq \frac{2E}{\chi^2}e^{-\sqrt{\frac{2}{\alpha'}}x^{0}}\ . 
\end{equation} 
This exponential fall-off agrees 
exactly with the string theory result from 
the boundary state \cite{Sen:2002in}. Note that this  prediction 
is different from that obtained using the BSFT 
effective action \cite{0205085}, where the 
square  of $x^{0}$ appears in the exponential. One can also see that
the BI-form $K=\sqrt{1+\kappa_{BI}(\partial T)^2}$ with the potential
\refs{VBSFT}
predicts that $(x^0)^2$ appears in the exponential.
In addition for $\chi<<1$ we
find the same dependence of $p$ and $E$ on $\chi$ 
as predicted from the boundary state \refs{p}.
On the other hand, 
in the boundary state approach, the pressure is always negative, whereas 
here  we find that the pressure approaches zero from above 
(see figure $2$). The same phenomenon  was 
also observed in \cite{0205085,0205098} for the BSFT effective 
action. Indeed it is clear from \refs{p} that the sign of the
pressure as $V\to 0$ is the opposite of  the sign of $K(-\dot T^2)$,
which is negative in BSFT and \refs{kinkyaction}, 
but positive for a BI-type action. 
There one of the  speculations 
was that this difference could be due to the fact that the solution of the 
BSFT 
differ from the exact marginal deformations. This possibility can be 
excluded here 
as the solutions to our action are precisely the marginal
deformations.
Although  higher derivative terms could lead to corrections in 
our action as well.

\begin {figure}
\center{\epsfig{file=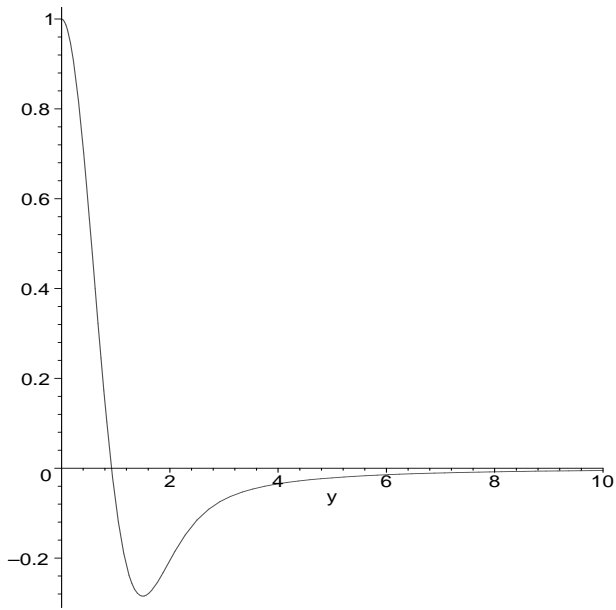,height=8cm,width=8cm}}
\caption{$\exp(-y^2)(\exp(y^2) + i\sqrt{\pi}y{\rm erf}(iy))$}
\end{figure}

Finally we make some comments on the interpretation of these
solutions. For the static solutions 
$T = \chi{\rm cos}(x^p/\sqrt{2\alpha'})$
the energy density is not spatially homogeneous but is peaked about
$x^p=\sqrt{\alpha'/2}\pi (2n+1) $ for integer $n$, becoming more sharply
peaked as $\chi \to \infty$. However, by construction the action \refs{KV} is 
independent of $\chi$ and therefore the total energy,
evaluated over a single period  $2\pi\sqrt{2\alpha'}$, 
is $4\pi\sqrt{\alpha'}\tau_p$ for all $\chi$. 
For $\chi=1/2$ this represents a configuration of BPS D$(p-1)$-branes
at each odd $n$ and anti-D$(p-1)$ branes at each even $n$. 
The interpretation of  other values of $\chi$ is less clear,
however these are no longer marginal deformations
once string loop corrections are considered \cite{Lambert:2000fn}.
Of course, the solutions involving  the marginal deformation 
$T = \chi{\rm sin}(x^p/\sqrt{2\alpha'})$ can be 
obtained from the previous ones by translation along $x^p$.

For time dependent solutions the energy density, $\epsilon$, is
spatially homogeneous. In particular for 
$T = \chi{\rm cosh}(x^0/\sqrt{2\alpha'})$ we have 
$\epsilon = \sqrt{2}{\tau_p}e^{-{\chi^2}/{\sqrt{2\alpha'}}}$, which is less than the false vacuum energy
density, while for 
$T = \chi{\rm sinh}(x^0/\sqrt{2\alpha'})$ the energy density
$\epsilon = \sqrt{2}{\tau_p}e^{{\chi^2}/{\sqrt{2\alpha'}}}$ 
is greater than the false vacuum energy density. 
These solutions are no longer related by temporal or spatial 
translation. Indeed while both solutions start and end in the vacuum as
$t \to \pm\infty$ the ${\rm cosh}(x^p/\sqrt{2\alpha'})$ solutions
never pass over the energy barrier at $T=0$ whereas the 
${\rm sinh}(x^p/\sqrt{2\alpha'})$ solutions travel from one vacuum to
the other. 
As we discussed in section 2,  in the full string theory  
there is a periodic dependence on 
$\chi$
for the $\chi{\rm cosh}(x^0/\sqrt{2\alpha'})$ solutions but not for the 
$\chi{\rm sinh}(x^0/\sqrt{2\alpha'})$ solutions.

\section{Coupling to Gravity} 

The relevance of scalar field action with higher than quadratic derivative 
terms for cosmology has been recognised a while ago 
\cite{Armendariz-Picon:1999rj},  
where it was argued that scalar field actions of the form 
$S=V(T)K((\partial T)^2)$ can produce inflationary scenarios 
(k-inflation) as well as late stage cosmological acceleration (k-essence). 
It is therefore interesting to analyse our tachyon action from this point 
of view. 
Related analysis were carried out in \cite{0205085,Gibbons,Padmanabhan:2002cp,Frolov:2002rr,Shiu,
Cline:2002it}
for the BI-type and  BSFT effective actions. 
We consider a general first derivative effective action of
the form \refs{ex1} minimally coupled to gravity
\begin{equation} 
{\cal L} = \sqrt{-g}\left( 
\frac{1}{2\kappa^2}R - V(T)K(\partial_\mu T\partial^\mu T)\right)\ . 
\end{equation} 
In $d=p+1$ dimensions and with 
$T$ assumed to be spatially homogeneous and time dependent the metric then 
takes the usual FRW form 
\begin{equation} 
ds^2 = -dt^2 + a(t)^2 ds_{d-1}^2\ , 
\end{equation} 
where $ds^2_{d-1}$ is a spatial manifold with  constant curvature $k$. 
For simplicity we consider the spatially flat case $k=0$. 
A convenient set of evolution equations is then simply (recall that
one of Einstein's equations is not independent)
\begin{equation}\label{mieo} 
H^{2}=\frac{\kappa^{2}}{(d-1)(d-2)}\epsilon 
\mtx{and} \dot\epsilon=-(d-1)H\left(\epsilon+p\right)\ , 
\end{equation} 
where $H\equiv\frac{\dot a}{a}$ is the Hubble constant and $\epsilon$ and $p$ 
are the energy density and pressure, respectively. 
If we now substitute our tachyon action these general formulae become 
\begin{eqnarray} 
H^2
&=&\frac{2\kappa^2}{(d-1)(d-2)}{V(\sqrt{T^2-2\alpha'\dot T^2})}\ ,
\nonumber\\ 
{d\over dt}\left( V(\sqrt{T^2-2\alpha'\dot T^2})\right)
&=&- (d-1)H\left( V(T)K(-\dot T^2)+V(\sqrt{T^2-2\alpha'\dot 
T^2})\right)\ .
\nonumber \\ 
\end{eqnarray} 
In particular at late times  $\dot T >> 1$  and we can approximate 
$VK \simeq -V(\sqrt{T^2-2\alpha'\dot T^2})/{2\dot T^2}$ so that  
$VK<<-V(\sqrt{T^2-2\alpha'\dot T^2})$. Hence we see that 
$V(\sqrt{T^2-2\alpha'\dot T^2}) = C^2a^{1-d}$ for a constant $C$ and
therefore
\begin{equation}\label{mex} 
a(t) \sim t^{\frac{2}{d-1}}\ . 
\end{equation} 
Thus the outcome of this analysis is identical with 
that obtained in \cite{Gibbons} 
for the BI action and describes matter dominated expansion. 
 
For intermediate times it is convenient to consider the master equation 
\begin{equation} 
\dot\epsilon=-\kappa\sqrt{\frac{d-1}{d-2}}\sqrt{\epsilon}\left(\epsilon+p\right)\ . 
\end{equation} 
From figure $2$ we then see that 
the evolution starts off with an inflationary 
phase $p=-\epsilon$ and then transforming smoothly into a matter 
dominated expansion 
\refs{mex} for late times. This qualitative behaviour is the same as found 
in \cite{0205085} for the BSFT action.

\section{Conclusion} 
 
In this paper we have discussed the general properties of first derivative 
tachyon effective actions.  For example  we showed that the 
asymptotic vanishing of the pressure for time-dependent tachyon profiles is 
relatively generic, although the details vary.
On the other hand we argued that the 
periodicity of the stress tensor under marginal deformations 
cannot be reproduced by any first derivative effective action. 
We also studied in detail the first derivative effective action that we
proposed in \cite{Lambert:2001fa} and showed that it reproduces many of the
expected features of tachyon dynamics, including several correct quantitative
features. However, it seems appropriate to mention the more pessimistic note 
that one could interpret the large discrepancies among the various proposed
effective actions for time dependent tachyons, compared with their 
striking similarity for static profiles, as an indication that 
the effective action approach will not be as successful in the time dependent
case. Indeed it has recently been observed from the boundary state 
that time dependent tachyons also couple 
to massive closed string fields with an exponentially increasing strength
\cite{Okuda:2002yd}, so that the truncation to low level string modes is
potentially artificial.

\section*{Acknowledgements} 
I.S. would like to acknowledge helpful discussions with A. Barvinski and 
A. Sen. N.D.L. would like to thank H. Liu for discussions and Trinity
College Dublin for its hospitality.

\end{document}